# Causal Digital Twin from Multi-channel IoT

## Dr. PG Madhavan

Seattle, USA June 2021
pg@jininnovation.com

**Abstract**: *Treating data from each channel/ sensor in an IoT installation on its own separately is wasteful. This article shows how to treat them as a multi-channel time series and introduces the **State-space model formulation of Structural Vector Autoregressive (SVAR) model and the use of time-varying Kalman Filter** for optimal estimation of causal parameters. "Ladder graphs" are introduced as a powerful visualization tool for SVAR estimates where both instantaneous and lagged causal factors are displayed and interactions analyzed.* **Ladder Graph IS the Causal Digital Twin (CDT);** *its use for multiple IoT applications that involve multi-channel time series are explored briefly.* **The main takeaway is that the NEXT STEP in IoT ML is the utilization of data from multiple sensors \*collectively\* as a single multi-channel time series**. *This article shows the way to do it and extract high-order (causal) information via our ladder graph based Causal Digital Twin!*

**All significant IoT installations** involve collecting data from multiple sensors connected to different assets (or parts of an asset). Treating data from each channel/ sensor on its own separately is a good starting point; however, significant information is unavailable if they are not treated together! *This article introduces the technically correct way to treat them TOGETHER as a multi-channel or vector time series – then, we can extract significantly useful information such as the causal effects among the various assets which is captured as a ladder graph, which we call "Causal Digital Twin".*

Causality is a hard concept but for the purposes of Internet of Things (IoT) engineering applications, we can stay clear of difficult topics such as, "*If everything is causal, is free-will possible*?"! First, a simple operational definition of Causality – *X "causes" Y means that changing X would change Y* (Woodward, 2003).

For "explainable" digital twins, we need the knowledge of cause-effect relationships; cause-effect relationship among different entities alone provides the invariant basis for the **explanation of the causal chain** that leads to a prediction or prescription. It is good to know "what will be" (prediction) but in addition, if we can know "what is possible AND impossible", our knowledge is more complete and we can safely prescribe actions to meet our objectives (in an industrial application, better productivity leading to better gross margins).

**Counterfactual statements** refer to what are possible or impossible, as opposed to what happens; it is through counterfactual experimentation on a Causal Graph that one can identify the perimeter of the solution space.

Causal Digital Twin (CDT) exposes **CAUSAL interactions** *and not just correlations among* connected assets. CDT models the DYNAMICS (as opposed to structural/ STATIC) aspects of interconnected assets. The MAIN use case is in engineering Operations and Production (and not maintenance).

The basics of CDT, Causal Graph and related formalities were introduced in two recent publications (Madhavan, April 2021 and Madhavan, May 2021) and will not be repeated here. In this article, I take **Structural Vector Autoregressive (SVAR)** model as the starting point.

This article is derivative of the excellent work of Hyvarinen, et al. (2010) and Chen at al. (2011). These two publications elucidate every subtle theoretical aspect of SVAR modeling as well multiple solutions (including connection to "blind signal separation", a non-obvious inverse model solution under certain assumptions of non-Gaussianity and non-stationarity). *I urge the reader to follow these two articles closely for technical clarifications.*

*What is original in this article is the **State-space model formulation of SVAR and the use of time-varying Kalman Filter** for optimal estimation of causal parameters. "Ladder graphs" are introduced as a powerful visualization tool for SVAR estimates where both instantaneous and lagged causal factors are displayed and interactions analyzed.*

## SVAR Model Basics:

SVAR models are used when working with multi-channel time series. Multi-channel time series is treated as a vector time series. Vector time series at each sampling instant are temporally correlated to past samples; also, if you take a snapshot at one time instant, there may be correlations among the elements of the vector. The former is modeled as a Vector Autoregressive (VAR) model. If we think of each channel as a "node" in a causal graph, we can call the causal factors among them as "self-node" and "inter-node" **lagged** causal factors. The latter "snapshot" model called Structural Equation Model (SEM) leads to "inter-node" **structural** causal factors. NOTE that there are NO self-node *structural* factors since a certain node cannot be the cause and effect; also, there can be NO "cycles" among *structural* nodes since one node cannot be the cause and effect through a chain of nodes instantly (within a snapshot).

Types of Causal Factors in SVAR:

1. Self-Node Lagged (SNL)
2. Inter-Node Lagged (INL)
3. Inter-Node Structural (INS)

Given a multi-channel time series in vector form, $\mathbf{y}_t$, we can write the Structural Equation Model (SEM) as follows.

$$\mathbf{y}_t = \mathbf{S}^0 \mathbf{y}_t + \mathbf{e}_t \qquad \ldots (1)$$

$\mathbf{S}^0$ is the matrix of INS causal factors with diagonal elements = 0 such that there is no self-causality.

Separately, we can write the traditional Vector Autoregressive (VAR) model as follows.

$$\mathbf{y}_t = \sum_{d=1}^{D} \mathbf{S}^d \mathbf{y}_{t-d} + \mathbf{e}_t \qquad \ldots (2)$$

Past values up to "D lags" may have impact the current $\mathbf{y}_t$. However, D=1 in many SVAR modeling for practical purposes. $\mathbf{S}^1$ matrix elements contain the SNL (along the diagonal) and INL causal factors.

Hyvarinen, et al. (2010) show a few simple theoretical examples which makes clear that one cannot estimate causal factors in equations (1) and (2) separately; INS factors can "leak" into SNL or INL! Therefore $\mathbf{S}^0$ and $\mathbf{S}^1$ have to be estimated jointly by forming the following equation (Chen, et al., 2011).

Generalized SVAR model –

$\mathbf{y}_t = \mathbf{S}^0 \mathbf{y}_t + \sum_{d=1}^{D} \mathbf{S}^d \mathbf{y}_{t-d} + \mathbf{B}\, \mathbf{e}_t$ ; $\mathbf{B}$ *diagonal with* $\{b_{ii}\}$ *such that* $\mathbf{e}_t \sim \mathcal{N}(\mathbf{0},\, \mathbf{I})$ . . . (3)

Chen et al. (2011) explains the need for $\mathbf{B}$ matrix and the important role that the free parameters in $\mathbf{B}$ plays in the overall estimation.

## Generalized SVAR State-space Model:

Arising from equation (3), a state-space model can be specified. The State equation and the measurement equation are as shown -

$\underline{\mathbf{s}}[n+1] = \underline{\mathbf{A}}\, \underline{\mathbf{s}}[n] + \underline{\mathbf{D}}\, \underline{\mathbf{q}}[n]$

$\underline{\mathbf{y}}[n] = \underline{\mathbf{H}}[n]\, \underline{\mathbf{s}}[n] + \underline{\mathbf{r}}[n]$   where $\underline{\mathbf{q}}[n] = \mathcal{N}[\mathbf{0}, \underline{\mathbf{Q}}(n)]$ and $\underline{\mathbf{r}}[n] = \mathcal{N}[\mathbf{0}, \underline{\mathbf{R}}(n)]$    . . . (4)

Here, $\underline{\mathbf{y}}[n]$ is the discrete-time counterpart of $\mathbf{y}_t$ in equation (3).

The work of Young (2011) has focused on time-varying estimation for the past many decades. Following his approach, different dynamics can be accommodated via structuring $\underline{\mathbf{A}}$ and $\underline{\mathbf{D}}$, properly (for example, as a block diagonal matrix).

$\underline{\mathbf{A}} = \begin{bmatrix} \alpha & \beta \\ 0 & \gamma \end{bmatrix}$ and $\underline{\mathbf{D}} = \begin{bmatrix} \delta & 0 \\ 0 & \varepsilon \end{bmatrix}$. The elements of $\underline{\mathbf{A}}$ and $\underline{\mathbf{D}}$ are called "hyper-parameters". They can be pre-selected or estimated optimally. We will pre-select them and hence $\underline{\mathbf{A}}$ & $\underline{\mathbf{D}}$ are not time-varying matrices themselves but they allow the States, $\underline{\mathbf{s}}$, to evolve in desirable ways. Choice of hyper-parameters leads to integrated random-walk process with its own characteristics.

To provide time-variability for the state space model in equation (4), each State is defined as a 2-component vector, the State's value AND its slope (which is the change in value over a sampling interval).

I.e., $\underline{\mathbf{s}}_i[n] = \begin{pmatrix} s_i[n] \\ \nabla s_i[n] \end{pmatrix}$ which in essence doubles the size of the State vector. What we get in return is the evolution of each State impacted by its slope also. Each State, 'i', evolves as follows:
$s_i[n] = \alpha\, s_i[n-1] + \beta\, \nabla s_i[n-1] + \delta\, q_i[n]$ and $\nabla s_i[n] = \gamma\, \nabla s_i[n-1] + \varepsilon\, q_i[n]$

Defining $\underline{\mathbf{A}}_O = \begin{bmatrix} \alpha & \beta \\ 0 & \gamma \end{bmatrix}$ and $\underline{\mathbf{D}}_O = \begin{bmatrix} \delta & 0 \\ 0 & \varepsilon \end{bmatrix}$,
the block-diagonal matrices (of appropriate dimensions) in equation (4) are –

$\underline{\mathbf{A}} = \begin{pmatrix} \mathbf{A}_O & 0 & 0 \\ 0 & - & 0 \\ 0 & 0 & \mathbf{A}_O \end{pmatrix}$ and $\underline{\mathbf{D}} = \begin{pmatrix} \mathbf{D}_O & 0 & 0 \\ 0 & - & 0 \\ 0 & 0 & \mathbf{D}_O \end{pmatrix}$

For the proper choice of $\underline{\mathbf{H}}[n]$ matrix, the State vector, $\underline{\mathbf{s}}[n]$, will converge to the elements of $\mathbf{S}^0$ and $\mathbf{S}^d$ in equation (3) which contains the Self-Node Lagged (SNL), Inter-Node Lagged (INL) and Inter-Node Structural (INS) causal factors. Going forward, D=1.

Any one channel of our multi-channel time series with "G" channels,

$y_k[n] = [y_1[n] \ldots \cancel{y_k[n]} \ldots y_G[n]\ y_1[n-1] \ldots y_k[n-1] \ldots y_G[n-1]] [s_{k1}[n]\ s_{k2}[n]\ \ldots\ s_{k,(G-1)}[n]\ s^D{}_{k1}[n]\ \ldots\ s^D{}_{kG}[n]]^T$

$y_k[n]$ on the right-hand side is crossed out since self-node structural causality does not exist as discussed. Gathering elements on the RHS into vectors, we can write $y_k[n]$ as -

$y_k[n] = \underline{Y}_k[n, n-1]\ \underline{S}_k[n]^T$

With $\underline{Z}$ as an appropriately dimensioned zero-vector, we can write the G channels of our time series as –

$$\begin{pmatrix} y_1[n] \\ \ldots \\ y_G[n] \end{pmatrix} = \begin{pmatrix} \underline{Y}_1[n, n-1] & \underline{Z} & \underline{Z} & \underline{Z} \\ & \ldots & & \\ \underline{Z} & \underline{Z} & \underline{Z} & \underline{Y}_G[n, n-1] \end{pmatrix} \begin{pmatrix} \underline{S}_1[n] \\ \ldots \\ \underline{S}_G[n] \end{pmatrix}$$

$$= \underline{H}[n]\ \underline{S}[n]$$

$\underline{H}[n]$ to be used in equation (4) for Kalman Filtering is thus obtained. $\underline{H}[n]$ is available at each 'n'; the only unknown is the State vector, $\underline{S}$.

Equation (4) can be solved optimally for States, $\underline{S}$, using the traditional Kalman Filter. Due to the augmented States, we obtain a time-varying estimate of $\underline{S}$ that is optimal for the linear case (note that Gaussian assumption is NOT necessary since it is not employed in the "innovations approach" to the derivation of Kalman Filter; Sarkka, 2013).

### Real Data Modeling: NASA Bearing data

We will develop Causal DT in a real-life setting using the popular NASA Prognostics Data Repository's bearing dataset. The data is from a run-to-failure test setup of bearings installed on a shaft. The arrangement is shown in figure 1.

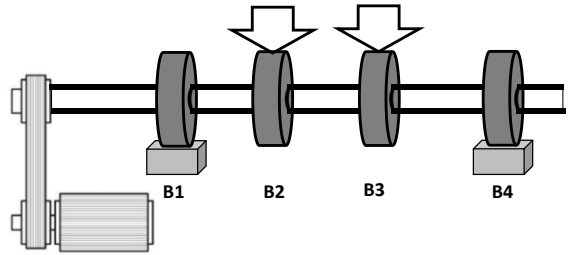

Figure 1. NASA Prognostic Data Repository Bearing data collection setup

Data were collected from Feb 12 to Feb 19, 2004 and the tests were run continuously to failure and data collected in blocks of 1 second through the entire period. Bearing 1 's outer race failed on Feb 19. The 4-channel time-series data of vibration measurements from each of the 4 bearings was subjected to our **Generalized SVAR state-space model and time-varying Kalman filter ("GSSM-TV Kalman filter")** to estimate the causal factors among these 4 bearings.

In an earlier study (Madhavan, April 2021), we had assumed that Causal Discovery was done already (by a machinery expert who provided us with the DAG corresponding to the 4-bearing system). The justification is that in IoT engineering applications, we are working with man-made equipment and systems and as such, *domain experts will be able to provide us a-priori with a very good approximation of the relevant DAG*. While we could have proceeded in the same manner, we want to assess the performance of GSSM-TV Kalman filter for BOTH causal discovery and causal estimation.

## NASA Bearing data results:

In our previous study (Madhavan, April 2021), once the Causal Estimation was done (using a different neural network-based method), we demonstrated its potential use for "what-if" analysis and "counterfactual" experiments – hence they will not be repeated here. We will simply focus on the causal factor estimation using GSSM-TV Kalman filter and its visualization via "ladder" graph here.

We took 1-second chunks of 4-channel data on each day from Feb 12 till Feb 19 when Bearing 1 failed. One can think of the results from Feb 12 as a baseline and watch the progression of Self-Node Lagged (SNL), Inter-Node Lagged (INL) and Inter-Node Structural (INS) causal factors.

Feb 18 data just prior to failure is the most interesting one. Figure 2 shows the 12 Structural and 16 Lagged causal factors estimated as States on Feb 18.

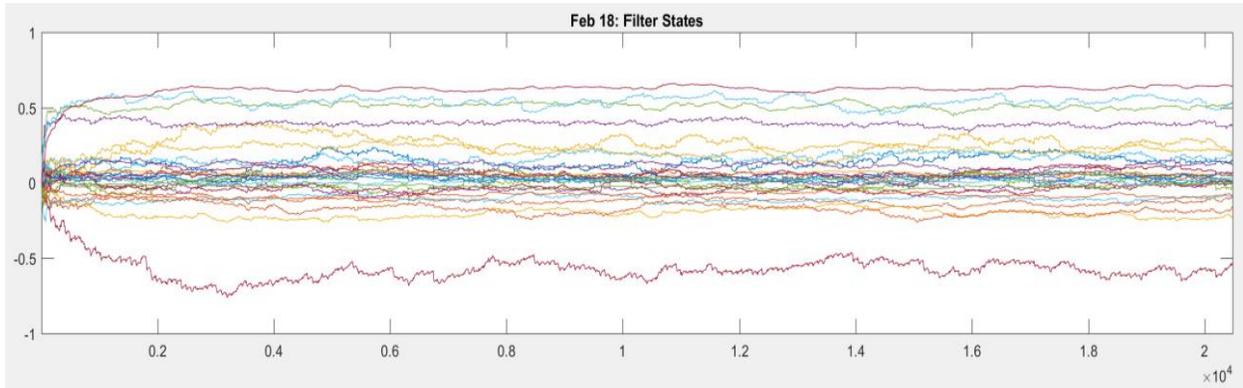

Figure 2. Trajectory of the States (causal factors)

Kalman filter converges quickly and reaches steady state after which there are some time-varying characteristics. The average of the last 5000 points of the 28 states are taken as the converged Self-Node Lagged (SNL), Inter-Node Lagged (INL) and Inter-Node Structural (INS) causal factors. Causal factors less than or equal to 0.1 were set to zero considering them negligible. The tables below show the INS factors on the left and SNL and INL factors on the right.

| B1 | B2 | B3 | B4 | |
|----|----|----|----|----|
| 0 | 0.1408 | 0 | 0.2626 | B1 |
| 0 | 0 | -0.1251 | -0.23 | B2 |
| 0 | -0.1855 | 0 | 0.226 | B3 |
| 0 | 0 | 0 | 0 | B4 |

| B1 | B2 | B3 | B4 | |
|----|----|----|----|----|
| 0.3926 | 0 | 0.1719 | -0.5922 | B1 |
| 0.1274 | 0.5031 | -0.1042 | 0 | B2 |
| 0 | 0 | 0.5469 | 0 | B3 |
| 0 | 0 | 0 | 0.6423 | B4 |

Instead of the usual graph diagram, we attempt to use a "ladder" graph as shown in figure 3. Differing from tradition, *our Ladder Graph permits non-horizontal links* also.

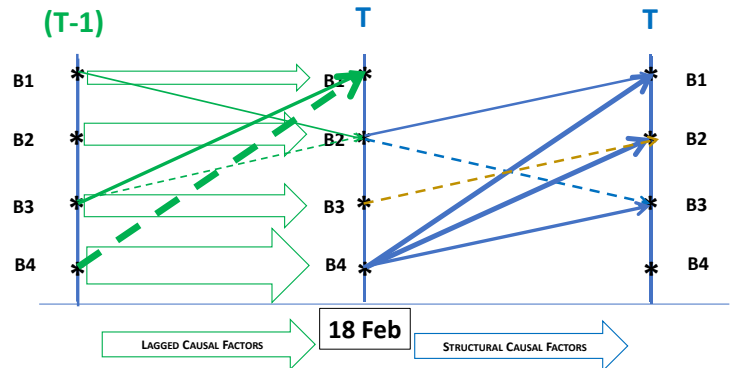

Figure 3. Ladder graph for Feb 18 causal factors

- The 3 vertical axes correspond to the past time instant (lag=1) and current time instant TWICE; lagged factors, SNL and INL, are shown going from (T-1) axis to (T) axis in green in the left pane. The right pane shows the structural INS causal factors in blue (T to T).
- The thickness of the arrow is proportional to the causal magnitude. Dashed arrows represent negative causal factors (meaning Cause will decrease the Effect).
- The block horizontal arrow in the left pane is a special case – they are the self-node lagged (SNL) causal factor which is nothing but the first AR coefficient; SNLs are very important in our ladder graph analysis.
- There are many geometric properties that can be extracted from such a ladder graph.
    - No horizontal arrows in the right panel allowed (= no self-node structural causality).
    - Horizontally aligned "X" indicates a closed loop which should not be present in a DAG; unfortunately, we do find the "X" pattern between B2 and B3 on the right-hand panel.
    - It is not clear at this point why there is such a degenerate case; it is possible that the factor magnitude of one of them was close to the rejection threshold but estimation errors caused its retention. Additionally, sampling rate can have an impact (sampling interval in NASA bearing case was 50 μ second) on this mislabeling.
    - In all the other days' ladder graphs, there is only one more such "X".

Keeping in mind that B1 failed on Feb 19, it is interesting to investigate whether the ladder graph on Feb 18 gives us any clues. A simplified ladder diagram compared to figure 3 for Feb 18 is shown in figure 4.

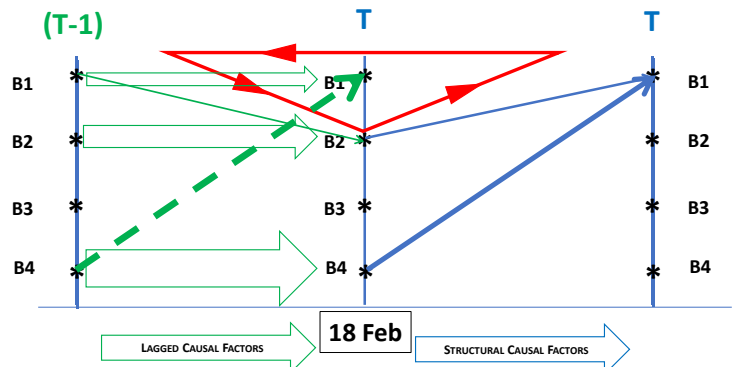

Figure 4. Simplified ladder diagram of Feb 18 for analysis

Following the causal factor effects from left to right, we can write the following statements/ propositions.

Note that what is on the right-most axis will be at the left-most axis *at the NEXT time instant*!

START at T=t
B1(t-1) ↑ B1(t)
B4(t-1) ↑ B1(t)

Next Instant, T=t+1
B1(t)   ↑ B1(t+1)        B1(t)   ↑ B2(t+1)        B2(t)   ↑ B2(t+1)
B4(t)   ↑ B1(t+1)        B2(t+1) ↑ B1(t+1)        B2(t+1) ↑ B1(t+1)

B4(t)   ↓ B1(t+1)
B4(t)   ↑ B4(t+1)
B4(t+1) ↑ B1(t+1)?

At the outset, note that in all 4 cases, B1 vibration INCREASES which is a precursor of its failure! B4 case is a bit unclear because it has a bi-directional causation effect on B1 at time, t+1, increasing and decreasing B1 vibration.

The upside-down isosceles triangle shown in red is another significant geometric signature – it is a positive feedback loop that will increase the vibration energy uncontrollably if left unchecked. In figure 3, you will notice a right-side up isosceles triangle B3-B2-B3 connected with dotted lines (negative causal factors) – this is a good loop in that it is negative feedback which will tend to reduce vibrations of B2 and B3. There may be other geometric objects in a ladder graph that has practical relevance yet to be discovered . . .

## Causal Digital Twins from GSSM-TV Kalman filter Ladder Graphs:

***At the simplest, Ladder Graph IS the Causal Digital Twin (CDT).*** Ladder graph as in figure 3 captures the entire dynamics of the connected system of "assets" (bearings in this case). Admittedly, on an IoT software UI, it can be displayed as overlays on the graphics (figure 1) of the physical system more attractively.

The propositions that we generated at T=t and (t+1) can be automatically generated by traversing all paths in the ladder diagram programmatically – *THIS is the most valuable outcome of this Causal Digital Twin*!

As is clear, GSSM-TV Kalman filter is a "learning" system that updates as each set of multi-channel time series data arrives. This update can be done in the Cloud (or at the Edge given proper CPU sizing) and UI can be made available at the Edge and on smartphones that Operations and Production staff carry.

The analysis of the Ladder Graph propositions along with all the "what-if" analysis and counterfactual experiments (see examples in Madhavan, April 2021) are typically performed only periodically, best done at a central planning/ assessment location using the Cloud infrastructure. As mentioned earlier, the main value proposition of CDT is its use in engineering Operations and Production that uses interconnected assets. However, some high-end condition monitoring (and predictive maintenance) can also be accomplished using the Ladder Graph CDT. Figure 5 shows the ladder graph on Feb 12, the day the system test was initiated when the bearing conditions are pristine.

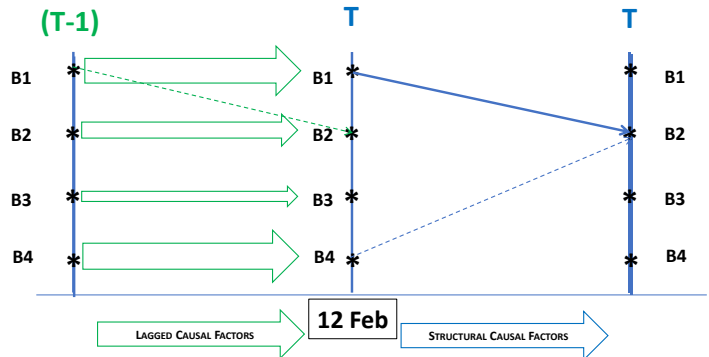

Figure 5. Start day of bearing test

There are only a few causal factors which are small in magnitude (thin lines); of course, self-node lagged causal factors are always present but you can see that there are no isosceles triangles or other positive feedback loops that creates run-away vibrations. When we inspect the ladder graphs in between the days from Feb 12 till Feb 18, we see a progression of new links (INL and INS) appearing and thickening. These observations can be translated into condition monitoring use on top of other simpler methods such as watching the amplitude of vibrations or simple ML methods based on single-channel time series data.

## Conclusions:

When IoT scenarios include an interconnected system of engineering assets, our **Generalized SVAR state-space model and time-varying Kalman filter ("GSSM-TV Kalman filter") based Ladder Graph CAUSAL Digital Twin** provides the system solution – a first of its kind as far as known.

Consider the following IoT use cases that generate multi-channel time series:

1. A manufacturing plant production line with a set of machines with sensors; OBJECTIVE: **Increase Production**
2. A building with monitoring data from HVAC system, occupancy, lights and computer operation, water usage; OBJECTIVE: **Minimize energy usage**
3. A retail store that monitors shelf occupancy, back-room store, shopper density, POS terminal data; OBJECTIVE: **Reduce OOS (out-of-stock) problem**
4. A smart city operation where multiple feeder road traffic and intersection traffic are monitored in real-time; OBJECTIVE: **Real-time traffic engineering to minimize congestion at the intersection**

*Wherever there is IoT data as multi-channel time series, our Causal Digital Twin can be applied for causal discovery and causal estimation to achieve some significant operational objectives as the above examples show.*

Currently, work has started as proofs-of-concept for the four IoT applications mentioned above and more are being added to establish the practical value of causal digital twins in improving productivity and quality as well as reducing waste, thus improving gross margins.

***The main takeaway is that the NEXT STEP in IoT ML is the utilization of data from multiple sensors *collectively* as a single multi-channel time series***. This article shows the way to do it and extract high-order (causal) information via our ladder graph based Causal Digital Twin!

**About the author:**

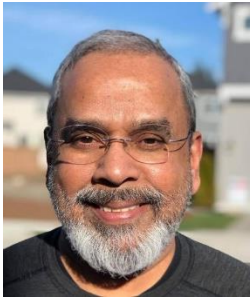

Dr. PG Madhavan launched his first IoT product at Rockwell Automation back in 2000 for predictive maintenance, an end-to-end solution including a display digital twin. Since then, he has been involved in the development of IoT technologies such as fault detection in jet engines at GE Aviation and causal digital twins to improve operational outcomes. His collected works in IoT is being published as a book, "Data Science for IoT Engineers" in June 2021. Rest of his career has been in industry spanning more major corporations (Microsoft, Lucent Bell Labs and NEC) and four startups (2 of which he founded and led as CEO). https://www.linkedin.com/in/pgmad/

**#Causality #Dynamics #Multichannel #Causaldigitaltwin #KalmanFilter #IoT #Laddergraph #Learning**